\documentclass[preprint,showpacs,preprintnumbers,amsmath,amssymb]{revtex4}
\usepackage{graphicx}
\usepackage{dcolumn}
\usepackage{bm}
\usepackage{latexsym}
\usepackage[dvips]{color}

\begin{document}

\markboth{Jun Goryo,Nobuki Maeda, and Ken-Ichiro Imura}{(De)confinement of supercurrent in $Z_2$ topological Insulators}

\title{(De)confinement of supercurrent in $Z_2$ topological Insulators}

\author{Jun Goryo$^1$, Nobuki Maeda$^2$, and Ken-Ichiro Imura$^3$}
\address{
$^1$Department of Physics, Nagoya University, Nagoya, 464-8602, Japan\footnote{Present Address: Institute for Industrial Science, The University of Tokyo, 4-6-1 Komaba, Meguro, Tokyo 153-8505} \\
$^2$Department of Physics, Hokkaido University, Sapporo, 060-0810, Japan \\
$^3$Department of Physics, Tohoku University, Sendai, 980-8578, Japan
}

\begin{abstract}
It is shown that a $Z_2$ topological system 
with $U_{\rm em}(1) \times U_z(1)$ (electromagnetic and spin) gauge symmetries shows 
superconductivity and quantized spin Hall effect simultaneously. 
When $U_z(1)$ is broken,  a dissipationless electric current is still 
possible to flow locally but net charge transfer is absent, i.e., {\it current is confined}. 
In the Kane-Mele model for graphene, this confining-deconfining transition of the supercurrent 
is driven by the Rashba spin-orbit interaction, which breaks $U_z(1)$.
\end{abstract}

\maketitle  



\section{Introduction} Dissipationless transport 
in a certain class of band insulators is often called ``topological".
Well-known examples are 
integer and fractional quantum Hall effects\cite{QHE}, 
specific type of anomalous Hall effect \cite{AHE}, etc. 
Dissipationless electric current in these examples is associated with time-reversal symmetry (TRS) breaking. 
However, topological transport for spin is not 
accompanied by breaking of TRS.
The Kane-Mele (KM) model for graphene\cite{Kane-Mele,Kane-Mele2} preserves TRS,
and realizes a quantized spin Hall (QSH) effect\cite{Kane-Mele2}.
Unfortunately, quantization of spin Hall conductivity (SHC) is not robust against perturbations 
which break the residual spin-rotation symmetry $U_z(1)$, 
e.g., Rashba spin-orbit interaction (SOI) $\lambda_R$ 
\cite{Haldane06}. 
In the presence of such perturbations, the role of spin Chern number,
characterizing the QSH state is 
replaced by a $Z_2$ topological invariant\cite{Kane-Mele,Fukui-Hatsugai}.
Physically, the same role is played by a 
pair of helical edge modes
\cite{Kane-Mele2,Bernevig}.
$Z_2$ classification of topological insulators is extended to three 
dimensions \cite{Fu-Kane,Fu-Kane2}.

$Z_2$ topological insulators show unique electromagnetic properties
\cite{Qi}.
In contrast to Ref. \cite{Qi}, which 
takes into account only
topologically {\it robust} quantities
protected by $Z_2$ invariants,
this paper
reveals the role of {\it non-robust} QSH effect in electromagnetic responses of
$Z_2$ topological insulator.
We show that QSH effect results in/from
massive electrodynamics characteristic to superconductivity. 
This mechanism of superconductivity is similar to the arguments in a TRS system \cite{BFSC} with ``topological order" characterized by the ground state degeneracy\cite{WenText}. 
On a torus geometry, the superconducting phase corresponds to {\it deconfinement} of
a dissipationless current (non-contractible current loops exist; see Fig.~\ref{torus}). Whereas a more generic $Z_2$ phase ($\lambda_R \ne 0$ case in KM model) corresponds to ``confining" state, where the current can flow locally  (contractible current loops exist; see, Fig. \ref{torus}), but no net current flows around the torus. Thus, we find that $\lambda_R$ is a  ``switch" for dissipationless charge transfer. 

Let us consider the following Hamiltonian of the KM model for graphene \cite{Kane-Mele,Kane-Mele2} in the continuum limit, 
$\hat{\cal{H}}=\hat{\cal{H}}_0 + \hat{\cal{H}}_\Delta + \hat{\cal{H}}_R$, which consists of the following three elements:
\begin{eqnarray}
\hat{\cal{H}}_0&=&v_F(i D_x \sigma_x \tau_z + i D_y \sigma_y),
\nonumber \\
\hat{\cal{H}}_\Delta&=&-\Delta \sigma_z \tau_z s_z, \
\nonumber\\
\hat{\cal{H}}_R&=&-{\lambda_R\over 2}(\sigma_x \tau_z s_y - \sigma_y s_x),
\label{KM}
\end{eqnarray}
where $D_x$ and $D_y$ are covariant derivatives defined below in Eq.~(\ref{Dmu}). 
Physical origins of each element composing $\hat{\cal{H}}$
are the following:
$\hat{\cal{H}}_0$ describes idealized graphene in the massless limit.
$\hat{\cal{H}}_\Delta$ gives a topological mass gap $\Delta$,
which stems from the intrinsic part of SOI.
$\hat{\cal{H}}_R$ is the Rashba term, induced by an electric field applied perpendicularly 
to the graphene layer, and breaks the spin-rotational symmetry, i.e., the conservation of $s_z$. 
Note that there are three species of (pseudo) spins in the model:
A-B sublattice spin $\vec{\sigma}$, valley (K-K') spin $\vec{\tau}$, 
and real spin $\vec{s}$, where, e.g.,
$\vec{\sigma}=(\sigma_x, \sigma_y, \sigma_z)$ represent Pauli matrices.

Our total Lagrangian density reads,
\begin{equation}
{\cal{L}}=\Psi^\dagger \left(i D_0 - \hat{{\cal{H}}}\right)\Psi +\frac{1}{2}(E^2 - B^2),
\label{L}
\end{equation}
in which we take into account the dynamics of electromagnetic fields (Maxwell term).
$\Psi$ represents 
the electron field.
The electromagnetic field $A_\mu$ $(\mu=0,\ x,\ y)$ couples to $\Psi$ 
through a covariant derivative $D_\mu$ in $\cal{L}$ defined by
\begin{eqnarray}
i D_\mu = i \partial_\mu - e A_\mu + {s_z\over 2} a_\mu.
\label{Dmu}
\end{eqnarray}
Here we have introduced a gauge field $a_\mu$,
which couples to the spin degrees of freedom of $\Psi$.
The {\it spin} gauge field $a_\mu$, thus introduced,
plays the central role in the following discussion.

Let us consider a functional integral with an additional term 
$a_\mu \bar{J}_{\rm sp}^\mu$
in Eq.~(\ref{L}), i.e.,   
\begin{eqnarray}
e^{i S_{\rm eff}[A^\mu,\bar{J}_{\rm sp}^\mu]}
\equiv 
\int {\cal{D}}\Psi^\dagger {\cal{D}}\Psi {\cal{D}} a_\mu e^{i \int d^3x ({\cal{L}}+ a_\mu \bar{J}_{\rm sp}^\mu)}. 
\label{FPI}
\end{eqnarray}
Since the 
(2+1)-dimensional ((2+1)D) spin current is defined by
$J_{sp}^\mu \equiv -\partial{{\cal{L}}}/\partial a_\mu$, 
the Lagrangian is rewritten as ${\cal{L}}={\cal{L}}_{a_\mu=0} - a_\mu J_{sp}^\mu$.
Then, integrating out $a_\mu$ in Eq. (\ref{FPI}), one finds
\begin{equation}
e^{i S_{\rm eff}[A^\mu,\bar{J}_{\rm sp}^{\mu}]}
=\int {\cal{D}}\Psi^\dagger {\cal{D}}\Psi \delta(J_{sp}^\mu - \bar{J}_{\rm sp}^\mu) e^{i \int d^3x {\cal{L}}_{a_\mu=0}},
\label{cstr}
\end{equation}
i.e., a constraint, $J_{\rm sp}^\mu=\bar{J}_{\rm sp}^\mu$,
is imposed on the spin current. 
Hence, 
$S_{\rm eff}[A^\mu,\bar{J}_{\rm sp}^{\mu}]$ 
is regarded 
as an effective action for the electromagnetic fields
under a given (2+1)D spin current $\bar{J}_{\rm sp}^\mu$. 
Eq.~(\ref{cstr}) thus clarifies the physical meaning of introducing the gauge field $a_\mu$.
Using gauge field is a standard prescription in field theory
for implementing some constraints on physical quantities \cite{WenText}. 
We will discuss elsewhere that we can also introduce $a_0$ and its integration to describe the electron correlation via 
Stratonovich-Hubbard transformation\cite{GM}. The result in 
the strong correlation limit is, in essence, equivalent to one shown in this paper.
 
In the following,
we derive the explicit form of $S_{\rm eff}[A^\mu,\bar{J}_{\rm sp}^\mu]$,
and discuss unique electromagnetic response properties of
the QSH insulator.
We consider, separately, (i) $s_z$-conserving ($\lambda_R=0$), and 
(ii) $s_z$-non-conserving ($\lambda_R\neq 0$) cases.
 
\section{$s_z$-conserving case: ``superconductivity" and QSH effect} 
When $\lambda_R=0$,
the system has a local $U_{\rm em}(1) \times U_z(1)$ symmetry, i.e.,
the Lagrangian density (\ref{L}) is invariant, independently, under 
$\Psi \rightarrow e^{-ie \theta} \Psi, A_\mu \rightarrow A_\mu + \partial_\mu \theta$, 
and under
$\Psi \rightarrow e^{i s_z \varphi / 2} \Psi, a_\mu \rightarrow a_\mu + \partial_\mu \varphi$.  
Integrating out $\Psi$ in Eq.~(\ref{FPI}),
\begin{eqnarray}
e^{i S_{\rm eff}[A^\mu,\bar{J}_{\rm sp}^\mu]}
=\int {\cal{D}} a_\mu e^{i S_{\rm eff}[A^\mu,a^\mu,\bar{J}_{\rm sp}^\mu]},
\label{SAa}
\end{eqnarray}
one finds,
in the Gaussian approximation, 
\begin{eqnarray}
S_{\rm eff}[A^\mu,a^\mu,\bar{J}_{\rm sp}^\mu]=
\int d^3x \left[\frac{\sigma_{xy}^s}{2} \epsilon^{\mu\rho\nu}(a_\mu \partial_\rho A_\nu + A_\mu \partial_\rho a_\nu) + 
\frac{\epsilon-1}{2}(e^2 - v_F^2 b^2) + \frac{\epsilon}{2}E^2 - \frac{1}{2\mu}B^2 + a_\mu \bar{J}_{\rm sp}^\mu\right]. 
\label{BF}
\end{eqnarray}
We have introduced the field strength, 
$e_i \equiv \partial_0 a_i - \partial_i a_0$, and $b \equiv \epsilon_{ij} \partial_i a_j$ ($i,j=x,y$),
associated with the gauge field $a^\mu$.
Note that the dielectric constant $\epsilon$ and the magnetic permeability $\mu$ are renormalized:
$\epsilon=1+1/ 6 \pi \Delta, \ \mu\equiv(1+ v_F^2 / 6 \pi \Delta)^{-1}.$
Here, we use the unit in which vacuum values of these constants are unity: $\epsilon_0=\mu_0=1$. 

The first terms ($a \wedge dA+A \wedge da$) on the {\it r.h.s.} of Eq.~(\ref{BF}) are
the so-called BF term \cite{BF}.
Like the Chern-Simons (CS) term ($A\wedge dA$, or $a\wedge da$)\cite{C-S},
these terms are topological,
i.e. independent of the metric of a manifold.
The CS term breaks TRS and is not induced in Eq.~(\ref{BF}). 
The BF term preserves TRS. Recall that TRS operation transforms
$A_0 \rightarrow A_0$, ${\bm A} \rightarrow - {\bm A}$, 
whereas, $a_0 \rightarrow -a_0$, ${\bm a} \rightarrow {\bm a}$. 

The coefficient in front of the BF term in Eq.~(\ref{BF}) is precisely 
the SHC $\sigma_{xy}^s$ 
(up to a symmetrization factor $1/2$); the latter obtained by Kubo formula. 
The physical impulication of BF term is, thus, the spin Hall effect. 
Note that SHC per a valley is $(e/4 \pi) {\rm sgn}(\Delta)$,  which is related to the winding number of Fermion propagator 
in (2+1)D momentum representation
\cite{C-S,Ishikawa-Matsuyama}. If one considers a Bloch system, 
i.e., a momentum space compactified on a two-torus,
one finds,  after frequency integration \cite{Ishikawa-Maeda-Ochiai-Suzuki},
$\sigma_{xy}^s=\frac{e}{2 \pi} C_{sc}$, where $C_{sc}$ is the spin Chern integer 
\cite{Haldane06} (See, also Refs. \cite{TKNN,Kohmoto85}). 
This quantization is assured whenever $s_z$ is conserved. 

We can see the bulk/edge correspondence as a consequence of the presence of $U_{\rm em}(1)\times U_z(1)$ gauge symmetry and quantum anomaly, which is an analogy of the field-theoretic argument of the edge state in quantum Hall system \cite{Wen} (see, also \cite{Edges}). The number of helical edge modes is counted by $C_{sc}$, equivalently by SHC $\sigma_{xy}^s$.  This fact has been already verified by Ref.~\cite{Haldane06} numerically. 

Note also that a kinetic term of (Maxwell term associated with) $a_\mu$ is also induced in Eq.~(\ref{BF}).
Its coefficient in front is $\epsilon-1$ (in contrast to $\epsilon$ for $A_\mu$), indicating that the term arises
due to the renormalization of dielectric constant.
The gauge field $a_\mu$ thus acquires its own dynamics. Some related arguments have been proposed \cite{Semenoff-Sodano-Wu}. 

The BF term gives the massive electrodynamics\cite{BF,BFSC}, analogous to the CS term\cite{C-S,anyonSC}. We proceed 
to integrate out $a_\mu$ in Eq.~(\ref{SAa}) using the Lorentz gauge.  
Then, we see that, for $\bar{J}^\mu_{\rm sp}=0$, 
\begin{eqnarray}
S_{\rm eff}[A^\mu,\bar{J}_{\rm sp}^\mu=0] =
\int d^3x \left[\frac{\sigma_{xy}^{s2}}{2(\epsilon-1)v_F^2}(A_0^{T2} - v_F^2 {\bm A}^{T2}) + \frac{\epsilon E^2 }{2}- \frac{B^2}{2\mu} \right].
\label{J=0}
\end{eqnarray}
The superscript $T$ denotes the transverse component:
$A_\mu^T=A_\mu - (\partial_\mu/\Box) \left[(1/v_F)^2 \partial_0 A_0 - \partial_j A_j\right]$,
where we introduced a d'Alembert operator (d'Alembertian) suitable for graphene:
$\Box\equiv (1/v_F)^2 \partial_0^2 - \partial_i^2$.
The presence of $A_0^{T2}$ and $A_i^{T2}$ terms in Eq. (\ref{J=0}) 
dictates that the electromagnetic field becomes indeed {\it massive}.  
These terms are known to 
be induced in the effective action for superconductivity,
after integrating out the Nambu-Goldstone mode \cite{Weinberg} (see also Ref.~\cite{Goryo-Ishikawa}). 

Note that (possible) superconductivity in topological insulators mentioned in Ref.~\cite{Grover-Senthil} is different from ours.
The former is predicted outside the QSH phase.
In our scenario, superconductivity arises in the QSH phase, and 
is shown to be indispensable for the occurrence of QSH effect.

Let us examine the QSH effect. 
In the presence of finite spin current,  $\bar{J}_{\rm sp}^\mu\neq0$, 
an additional term,
$\frac{\sigma_{xy}^s}{\epsilon-1} A^\mu \frac{\epsilon_{\mu\rho\nu}\partial^\rho}{{\Box}}  \bar{J}_{\rm sp}^\nu 
+ {\cal{O}}(\bar{J}^{\mu 2}_{\rm sp})
$, is in the effective action (\ref{J=0}). 
Suppose that we apply a static spin current in the $y$-direction:
$\bar{J}_{\rm sp}^y=\bar{J}_{\rm sp}^y(x)$, $\bar{J}_{\rm sp}^0=\bar{J}_{\rm sp}^x=0$. 
Then, the equation of motion for ${\bm E}$ reads, 
\begin{equation}
\frac{d^2}{d x^2} \epsilon E_x(x)=
\frac{\sigma_{xy}^{s2}}{\epsilon-1} E_x(x) + \frac{\sigma_{xy}^s}{\epsilon - 1} \bar{J}_{\rm sp}^y(x).
\label{QSH}
\end{equation}
In the long-wavelength limit,
the second order derivative term on the {\it l.h.s.} of Eq.~(\ref{QSH})
is irrelevant, leading to the QSH effect.
The presence of $\sigma_{xy}^{s2}$ term coming from the gauge mass term is essential. 
The equation of motion for the magnetic field ${\bm B}$
is analogous to the one for a thin-film superconductor \cite{Pearl}.

\section{$s_z$-non-conserving case: confinement of dissipationless
current }  When $\lambda_R \neq 0$, the SHC is not quantized, 
and loses its topological meaning \cite{Haldane06}. However, it still plays an important role for the electromagnetic response.
The action $S_{\rm eff}[A^\mu,a^\mu,\bar{J}_{\rm sp}^\mu]$ in Eq.~(\ref{BF}) obtained 
after fermion integration is modified and a mass term of $a_\mu$,  $\frac{m_a^2}{2} (a_0^2 - v_F^2 a_i^2)$, appears 
as a direct consequence of the breakdown of local $U_z(1)$ symmetry. Here, 
$m_a^2=(\lambda_R^2/8 \pi |\Delta|) + {\cal{O}}(\lambda_R^4)$.
SHC, and other coefficients $\epsilon$ and $\mu$ in  Eq.~(\ref{BF}) 
receive corrections in the order of ${\cal{O}}(\lambda_R^2)$. Thus, SHC is no longer quantized.  
By integrating out $a_\mu$ for $\bar{J}^\mu_{\rm sp}=0$, we obtain  
\begin{eqnarray}
S^{\lambda_R \neq 0}_{\rm eff}[A^\mu,\bar{J}_{\rm sp}^\mu=0]=
\int d^3x \left[
\frac{\sigma_{xy}^{s 2}}{2(\epsilon-1)v_F^2}
\left(A_0^T\frac{{\Box}}{{\Box}+m_a^2}A_0^T- {\bm A}^T\frac{v_F^2 {\Box}}{{\Box}+m_a^2}{\bm A}^T\right)  
+\frac{\epsilon E^2 }{2}- \frac{B^2}{2\mu} \right]. 
\label{Seff2}
\end{eqnarray}
If we introduce $\bar{{\bm J}}_{\rm sp}\neq0$, 
the spin Hall effect occurs with non-quantized SHC. 

\begin{figure}
\begin{center}
\includegraphics[width=7cm]{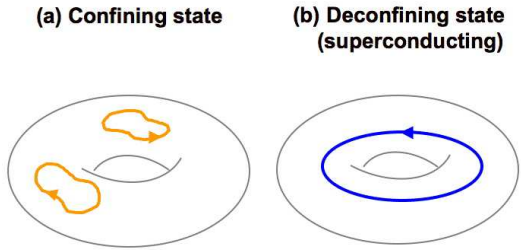}
\caption{\label{torus} 
Schematics of possible current loops in 
(a) confining state, and (b) deconfining state for dissipationless electric current.}
\end{center}
\end{figure}

We first introduce the notion of {\it confinement} of dissipationless current. 
We consider a system on a two-torus to avoid unnecessary complications due to the
presence of an edge. 
The Fourier transform of static charge current density in Eq.~(\ref{Seff2}) is 
\begin{eqnarray}
{\bf j}_{\bm q}=-\frac{\sigma_{xy}^{s 2}}{\epsilon - 1}\frac{{\bm q}^2}{{\bm q}^2 + m_a^2} {\bm A}_{\bm q}, 
\label{jq} 
\end{eqnarray}
which is dissipationless, since it is driven by a static vector potential 
that never yields a voltage drop. 
Eq.~(\ref{jq}) also indicates that current density ${\bm j} ({\bm x})$ can be finite, on the other hand, the net current at an arbitrary cross section of the torus is zero, i.e., current is confined\footnote{Color charge of quarks is confined and meson and baryon must be color singlet in the quantum chromodynamics\cite{Kogut}. It is rather intuitive, but the fact that total current must be zero remind us the color confinement when we see an analog between non-zero local current density and color of a quark. }. 
The latter can be shown by using 
the current conservation relation and the fact that ${\bm j}_{{\bm q}=0}=0$. We note that  ${\bm j}_{{\bm q}=0}$ is equal to the integral of current density over the entire surface of the torus. 
Absence of net current is a usual insulating behavior. 
Unusual fact in this confining state is that a finite current can flow locally. 
Such a local current circulates around a contractible closed contour (see Fig. \ref{torus} (a)).

\section{Deconfining (superconducting) phase revisited} We have seen 
that spin-rotation symmetry breaking yields the confinement of dissipationless charge current. 
On the other hand, in the spin-rotation symmetric limit, $m_a \rightarrow 0$, Eq.~(\ref{jq}) reduces to,
${\bm j}_{\bm q}=-{\sigma_{xy}^{s2} \over \epsilon - 1} {\bm A}_{\bm q}$,
i.e., identical to
the supercurrent obtained from Eq.~(\ref{J=0}).  
This means that ${\bm j}_{{\bm q}=0}$ can be finite, 
allowing for a finite net current circulating around the torus, 
i.e., ``deconfinement" is accomplished. 

In the deconfining state, non-contractible current loops become possible (see Fig.~\ref{torus} (b)).
On the other hand,  contractible loops are forbidden because of the superconducting nature of the state 
(perfect diamagnetism).
This contrasting behavior is summarized in Table~\ref{table}.
The topology of possible current loops thus plays the role of ``order parameter" which classifies 
the confining and deconfining phases. This fact reminds us the role of Wilson loop in the lattice gauge field theory\cite{Kogut}. 
\begin{table}
\caption{Topology of possible current loops on the torus.}
\begin{center}
\begin{tabular}{c|c|c}
\hline\hline
&\multicolumn{2}{c}{current flow along a}\\
& contractible loop & non-contractible loop \\
\hline\hline
QSH phase & NO & YES\\
($\lambda_R=0$)&&$\rightarrow$ deconfining phase \\
\hline
$Z_2$ phase & YES &  NO  \\
($\lambda_R\ne 0$) & &$\rightarrow$ confining phase\\ 
\hline\hline
\end{tabular}
\end{center}
\label{table}
\end{table}

This confinement-deconfinement transition is expected in graphene 
as the Rashba coupling $\lambda_R$ is switched on 
by an electric field perpendicular to the system. It would be interesting 
if the transition can be applied to the development of switching devices for dissipationless charge transfer. It should be noted that confinement occurs even for the infinitesimally small spin-rotation symmetry breaking (see Eq.~(\ref{jq})). 
Then, a perfect-spin-conserving system would be needed to observe the deconfining (superconducting) state. 

\section{HgTe/CdTe quantum well}
Although we have used the 
KM model for demonstration, 
the basic ideas presented in this paper will be applicable
to other QSH insulators. The present result with spin-rotation symmetry would be relevant to 
HgTe/CdTe quantum well \cite{Bernevig,Morenkampf}, which has a large band gap ($\sim$100K) in bulk spectrum assisting the experimental observation. 
In this system, we can define a conserved pseudo spin associated with the band index, although real spin is not conserved. In parallel, we can see that the superconductivity is induced by integrating out Fermion and pseudo spin gauge field. Although protected from the thermal activations by the large band gap, this superconductivity is fragile against the pseudo spin-rotation symmetry breaking perturbation.  

\section{Summary}

It is shown that the KM model in $s_z$-conserving case shows 
superconductivity and quantized spin Hall effect simultaneously. 
When $s_z$-conservation is broken by the Rashba spin-orbit interaction,  
a dissipationless electric current is still possible to flow locally but net charge transfer 
is absent, i.e., {\it current is confined}. 

We will discuss elsewhere that we can also introduce the $0$-th component of the spin gauge field and its integration to describe the electron correlation via Stratonovich-Hubbard transformation. The result in 
the strong correlation limit is essentially equivalent to one which has been shown in this paper.

The authors are grateful to D. S. Hirashima and A. Tanaka for their useful comments. 
This work is supported by Grant-in-Aid for Scientific Research (No. 19740241) from the Ministry of Education, Culture, 
Sports, Science and Technology.

\end{document}